\documentclass{article} 
\usepackage{nips13submit_e,times}
\usepackage{url}

\usepackage{amsmath}
\usepackage{amsthm}
\usepackage{thmtools}
\usepackage{amsmath}
\usepackage{amsfonts}
\usepackage{paralist}
\usepackage{graphicx}
\usepackage{caption}
\usepackage{subcaption}
\usepackage{array}
\usepackage[numbers, sort&compress]{natbib}

\title{Refining the Semantics of Social Influence}

\author{
Katerina Marazopoulou\\
School of Computer Science\\
University of Massachusetts Amherst\\
Amherst MA, 01002 \\
\texttt{kmarazo@cs.umass.edu} \\
\And
David Arbour\\
School of Computer Science\\
University of Massachusetts Amherst\\
Amherst MA, 01002 \\
\texttt{darbour@cs.umass.edu} \\
\And
David Jensen\\
School of Computer Science\\
University of Massachusetts Amherst\\
Amherst MA, 01002 \\
\texttt{jensen@cs.umass.edu} 
}

\nipsfinalcopy 

\theoremstyle{definition}

\begin{document}

\maketitle

\begin{abstract}
\begin{quote}
With the proliferation of network data, researchers are increasingly focusing on questions investigating phenomena occurring on networks. 
This often includes analysis of peer-effects, i.e., how the connections of an individual affect that individual's behavior.
This type of influence is not limited to direct connections of an individual (such as friends), but also to individuals that are connected through longer paths (for example, friends of friends, or friends of friends of friends).
In this work, we identify an ambiguity in the definition of what constitutes the \emph{extended neighborhood} of an individual.
This ambiguity gives rise to different semantics and supports different types of underlying phenomena.
We present experimental results, both on synthetic and real networks, that quantify differences among the sets of extended neighbors under different semantics.
Finally, we provide experimental evidence that demonstrates how the use of different semantics affects model selection.
\end{quote}
\end{abstract}

\section{Introduction}
The growth of the internet has given rise to a plethora of online social systems that capture social and professional interactions among people, including Facebook, Twitter, LinkedIn, and many more.
Such systems produce large amounts of data that record the attributes of individuals, but also information about the connections and interactions among those individuals. 
These rich data sets allow researchers to investigate various aspects of social behavior on a scale that previously was not possible. 
Many of these phenomena involve \emph{social} or \emph{peer-effects} which measure the level of influence exerted on individuals by their \emph{neighborhood}---other individuals that are near to them in the network. 
Surprisingly, no clear semantics exist for defining the set of nodes that belong to that neighborhood. 
As we show in this paper, ambiguity in how neighborhoods are defined can have significant impact on analyses examining these effects.
Specifically, values for the aggregates of these sets can differ substantially depending on the specific semantics used, which may lead to different conclusions.
This is especially important considering the fact that large online studies often deem very small effect sizes to be significant.
For example, in one study \cite{kramer2014experimental} an effect size of $10e^{-3}$ was declared to have a significant impact because it translates to tens of thousands of users for a large network such as Facebook.

\begin{figure}
\centering
\begin{subfigure}[t]{.45\textwidth}
\centering
\includegraphics[width=.45\textwidth]{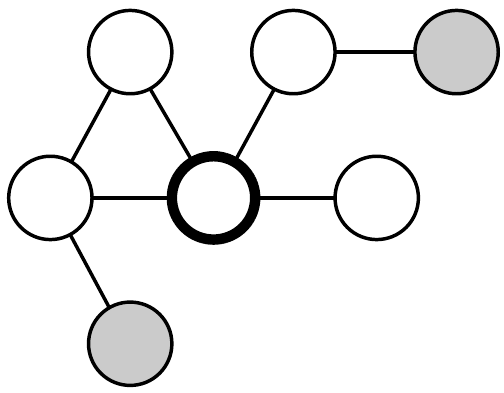}
\caption{Friends of my friends who are not my friends.}
\label{fig:bb}
\end{subfigure}\quad
\begin{subfigure}[t]{.45\textwidth}
\centering
\includegraphics[width=.45\textwidth]{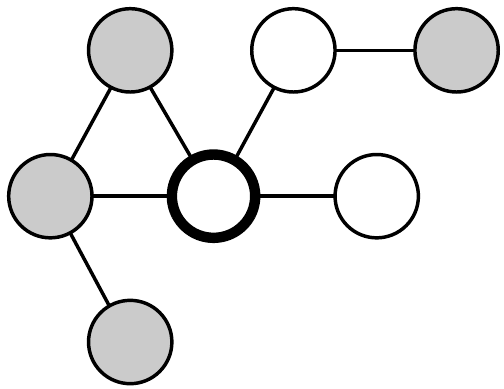}
\caption{Friends of my friends who could be my friends.}
\label{fig:nbb}
\end{subfigure}
\caption{Example of alternative semantics on a small network. The node with the highlighted outline is the central node of the analysis. Grey nodes are the ones that belong to the set of ``friends of friends'' for the central node under different types of semantics.}
\label{fig:alternative-semantics}
\end{figure}

As a motivating example, consider the following question:
\emph{does the happiness of an individual depend on the happiness of her extended social circle (friends of friends)?}
On the face of it, this problem is relatively straightforward to formulate. 
First, we would gather data about an individual's extended social circle. 
We can then aggregate those values and estimate if they have an effect on her happiness\footnote{Note that we are intentionally ambiguous regarding the exact method of estimation. 
The implications of this work apply to a large number of techniques both predictive and causal.}. 
However, we have failed to be explicit about a critical component: \emph{what exactly constitutes membership in someone's extended social circle?}
There are (at least) two equally reasonable options:
\begin{enumerate}
\item Only include individuals who are connected to one of her friends but not to her (friends of her friends who are not her friends).
\item Include all individuals connected to one of her friends, including those that are her friends (friends of her friends no matter whether they are her friends or not). 
\end{enumerate}
Figure~\ref{fig:alternative-semantics} provides a visualization of these different semantics for a given node of a small network.
While both options are reasonable, they lead to significantly different estimates of the values of the set, as we show in this paper.

Intuitively, those semantics capture different phenomena.
The first case captures phenomena where each node can be characterized by the length of the shortest path from that node to the central node of the analysis. 
That includes, for example, diffusion of information in networks. 
The second type, focuses on capturing properties of the extended neighborhood of a node.

The effects of using different semantics are amplified when considering longer paths of dependence (e.g., friends of my friends of my friends, etc.), since longer paths admit more semantics. 
Such longer-range dependencies are not unrealistic. 
In fact, according to Christakis and Fowler~\cite{christakis2009connected}, social networks obey the ``three-degrees of influence" rule. 
In other words, an individual can affect others that are up to three connections away (friends, friends of friends, friends of friends of friends).

In what follows, we provide a formal description of the problem and we define two different semantics.
We then provide experimental evidence using the Erd\"{o}s-R\'{e}nyi model, Barab\'{a}si model, and Watts-Strogatz model that quantifies the effect of using the different semantics on both set construction and subsequent model-selection. 
Finally, we examine the impact of these semantics on sixteen real-world networks from the Stanford network analysis project (SNAP) repository.

\section{Problem Description}

In this work, we consider attributed single-entity networks\footnotemark{}.
More formally, an attributed graph $G=(V, E, \mathbf{W})$ consists of a set of vertices $V$, a set of edges $E\subseteq V\times V$, and a $n\times |V|$ matrix of attribute values.
Each vertex is labeled with $n$ attributes and the value of the $i$-th attribute for vertex $v$ is given by $\mathbf{W}[i, v]$.
We restrict our attention to undirected attributed graphs, but the main principles easily extend to directed graphs. 
\footnotetext{Future work will address the more general setting of multi-entity networks.}

Here our focus is on peer or social effects, specifically, on the effect of the friends of your friends on you. 
Assume that we are given an initial vertex, $v_0\in V$. 
Informally, we can think of $v_0$ as the individual of interest.
In this work, we investigate the effect of two different semantics for constructing the set of friends of friends of $v_0$.
Let $V_k$ be the set of vertices that can be reached from $v_0$ with a path length of $k$ (without repetition of edges along that path).
One construction of our set is to simply consider all vertices that are found at path length 2, i.e., $V_2$.
Another approach is to include only the vertices whose shortest path is of length two, i.e. $V_2 \setminus V_1$.
The question that remains is what is the effective difference between employing those two strategies in terms of covariate estimation and model selection. 
This is our focus for the remainder of the paper. 

\section{Synthetic Experiments}

\subsection{How different are the sets constructed using the different semantics?}
The simplest measure to capture the difference between these semantics is to quantify the disparity between the sets constructed using the two strategies described above. 
Note that we are ignoring attribute values and focusing on the simple presence/absence of a vertex in a set. 
In this experiment, we first generated random networks with varying size (10, 50, 100, 200, 300, 400, 500 nodes)
using the following models:
\begin{enumerate}
\item Erd{\"o}s-R\'{e}nyi model~\cite{erdos1959random}. Specifically, we used the $G(n, p)$ model, which generates a random graph with $n$ nodes and any two nodes are connected with probability $p$.
For this experiment we used $p=$0.1-0.9 incremented by intervals of 0.1.
\item Barab\'{a}si-Albert model~\cite{barabasi1999emergence}. This model creates networks starting with one vertex and adding new vertices at each time step that are connected to existing vertices with probability proportional to the number of links that a vertex has (preferential attachment). 
For this experiment, we varied the power of the preferential attachment (0-3 by increments of 0.5).
\item Watts-Strogatz model~\cite{watts1998collective}. This model starts with a regular ring lattice with $n$ nodes, each connected to $k$ neighbors on both side. Then every edge is ``rewired'' with probability $p$ to connect to a different node (avoiding self-loops).
For this experiment, we varied the size of the neighborhood (1, 5, 10) and the rewiring probability (0.1-0.9 by increments of 0.2).
\end{enumerate}

\begin{figure}[t]
\centering
\includegraphics[width=0.7\textwidth]{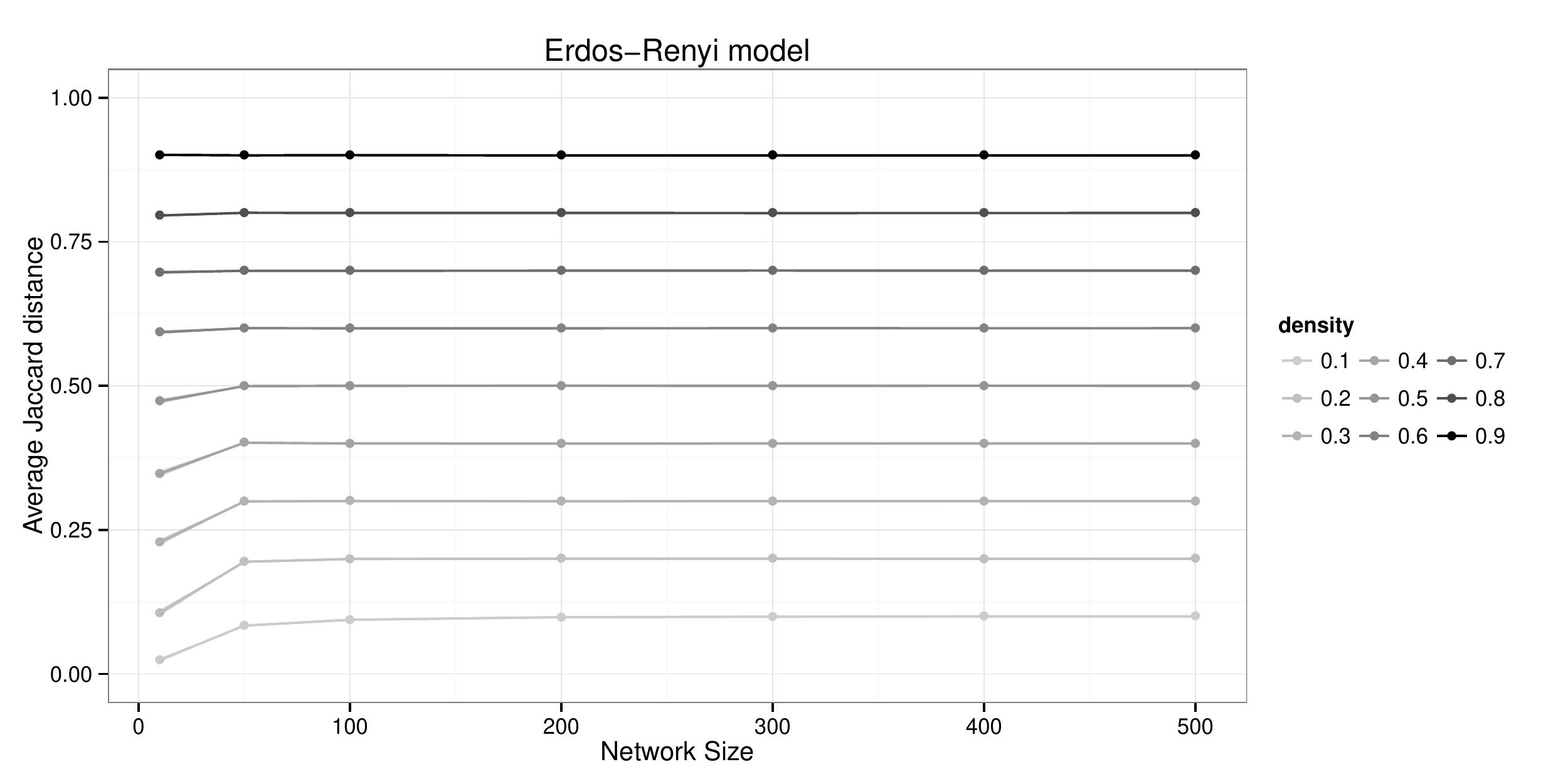}
\caption{Average Jaccard distance for the two different semantics on random networks generated using the Erd\"{o}s-R\'{e}nyi model with varying $p$ and network size.}
\label{fig:freq_erdos}
\end{figure}

\begin{figure}[t]
\centering
\begin{subfigure}[t]{.3\textwidth}
\includegraphics[width=\textwidth]{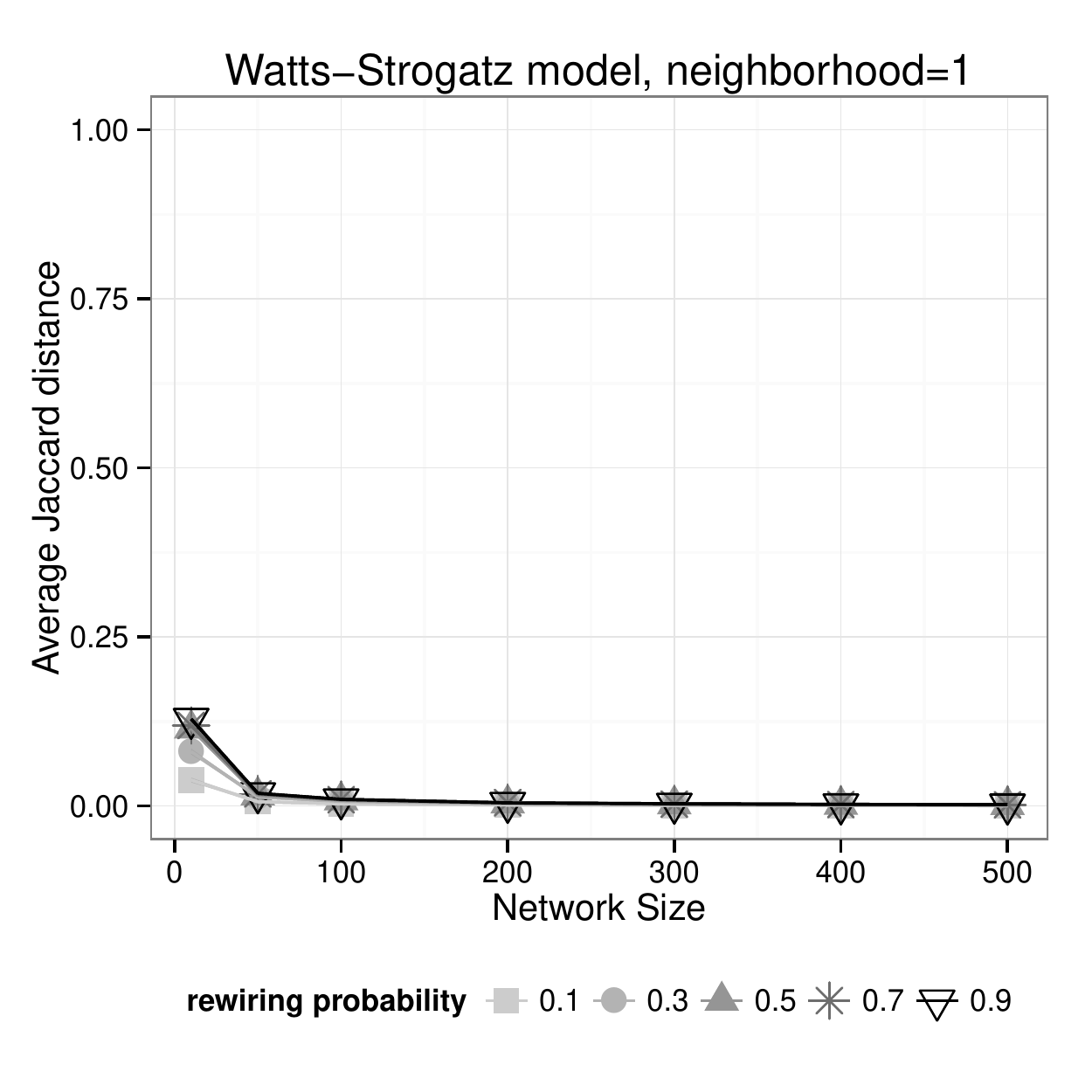}
\caption{neighborhood=1}
\label{fig:freq_sw1}
\end{subfigure}
\quad
\begin{subfigure}[t]{.3\textwidth}
\includegraphics[width=\textwidth]{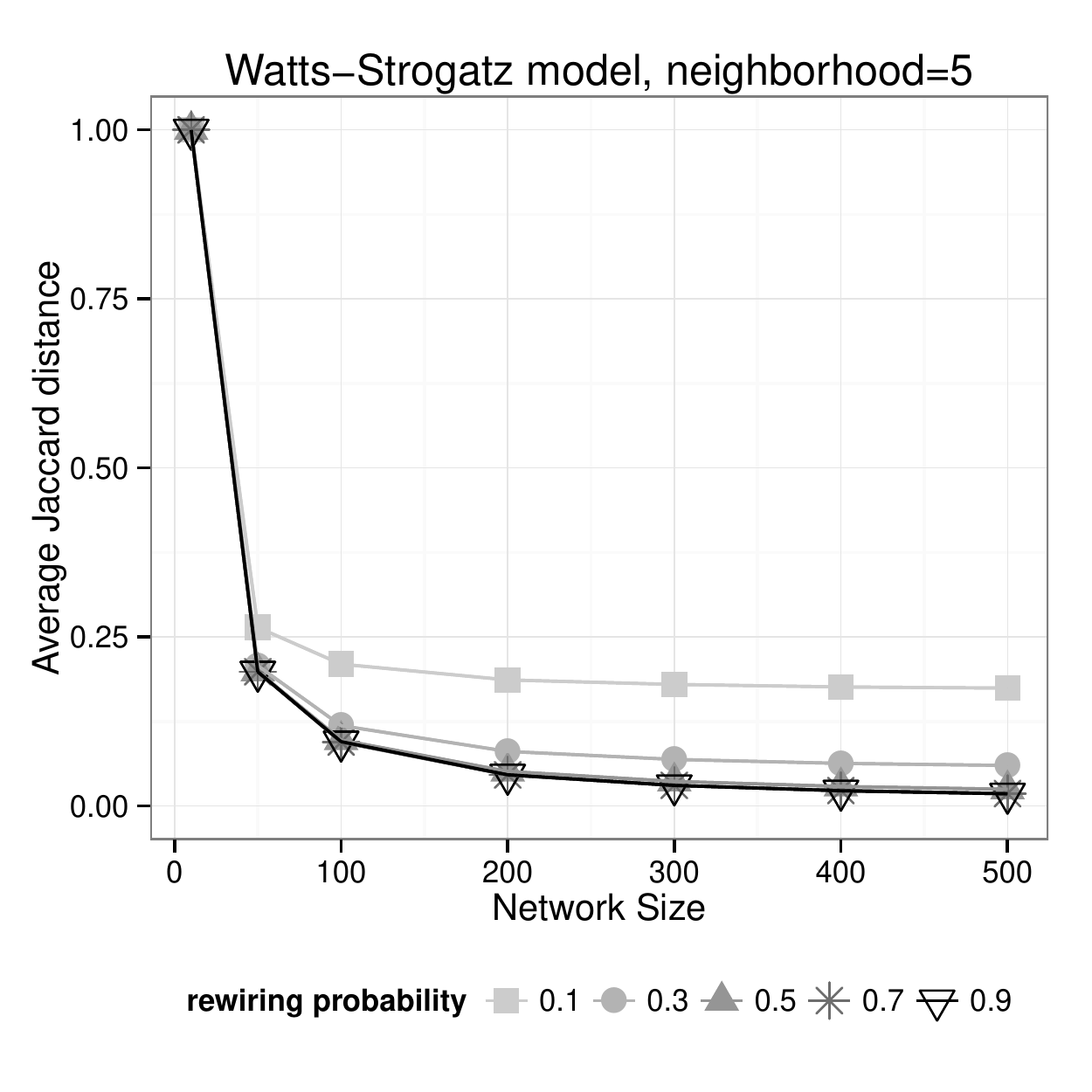}
\caption{neighborhood=5}
\label{fig:freq_sw2}
\end{subfigure}
\quad
\begin{subfigure}[t]{.3\textwidth}
\includegraphics[width=\textwidth]{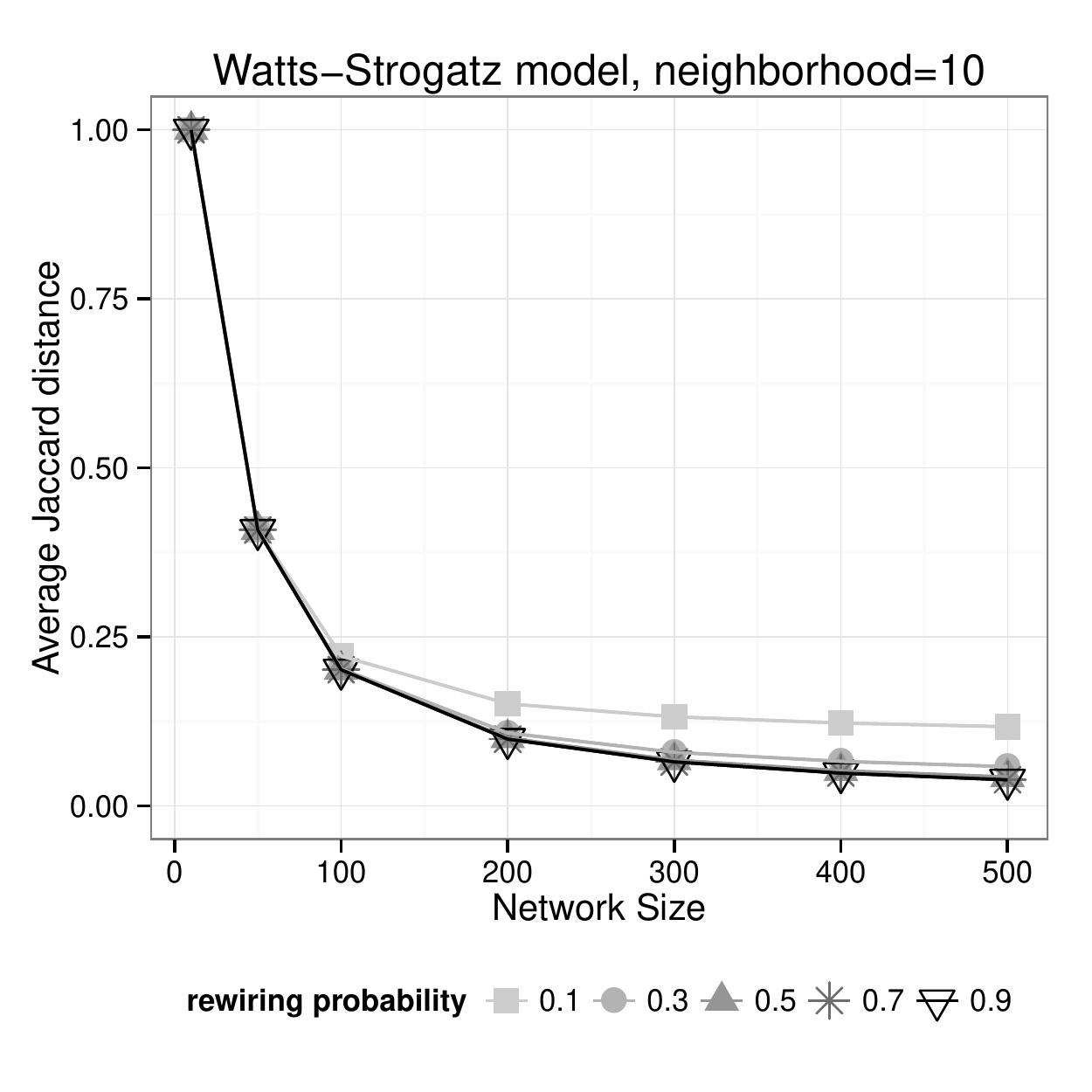}
\caption{neighborhood=10}
\label{fig:freq_sw3}
\end{subfigure}
\caption{Average Jaccard distance for the two different semantics on random networks generated using the Watts-Strogatz model with varying neighborhood size, rewiring probability, and network size.}
\label{fig:freq_sw}
\end{figure}

We then measured the Jaccard distance between the sets constructed under the different semantics, averaged across all nodes of a network and averaged over 500 trials.
The Jaccard distance between two sets $A$ and $B$ is defined as $J(A, B) = 1 - \frac{|A \cap B|}{|A \cup B|}$.
Intuitively, this quantifies the overlap of two sets, while accounting for the size of both.
For networks generated using the Barab\'{a}si model, the two semantics lead invariably to the same sets (Jaccard distance is 0).
The results for networks generated by Erd{\"o}s-R\'{e}nyi models are shown in Figure~\ref{fig:freq_erdos}.
As the density of the network increases (higher $p$), the Jaccard distance increases as well.
Finally, the results for networks generated using the Watts-Strogatz model are shown in Figure~\ref{fig:freq_sw}.
For values of the neighborhood parameter larger than 1, the two different semantics result in sets that have a non-zero Jaccard distance.

\subsection{How does parameterizing w.r.t.\ different semantics affect fit of the model?}

This experiment goes one step beyond measuring the distance between the sets created under different semantics. 
We seek to assess how the different semantics affect model selection.
We consider attributed graphs, where each node has two attributes, treatment $T$ and outcome $O$. We assume that the true generating model has a dependency of the form:
``The treatment $T$ of the friends of my friends affects my outcome $O$.''
We generated synthetic datasets using the following procedure.
\begin{itemize}
\item We first generated network structures using the Erd{\"o}s-R\'{e}nyi model and the Watts-Strogatz model.
For this part, we either varied the size of the network, or the parameters of the generating model (probability $p$ for the Erd{\"os}-R\'{e}nyi model, and neighborhood and rewiring probability for the Watts-Strogatz model).
\item We generated values for the attributes $T$ by drawing from a normal distribution $\mathcal{N}(\mu,\sigma)$, where $\mu\sim\mathcal{U}(-5, 5)$ and $\sigma\sim\mathcal{U}(0,3)$.
\item We generated values for $O$ using the conditional distribution $O\sim\mathit{agg}(\mathit{FriendsOfFriends}) + \epsilon * \mathcal{N}(0,1)$, where $\epsilon$ is the noise coefficient, $\mathit{FriendsOfFriends}$ is the set of $X$ values for the nodes that belong to the set of friends of friends (defined using two different semantics), and $\mathit{agg}$ is an aggregation function applied to the values of that set. 
For this part, we ran two sets of experiments, one using $\mathit{mean}$ as the aggregation function (i.e., the mean of the set of values in the sets $\mathit{FriendsOfFriends}$), and one using $\mathit{degree}$ (i.e., the size of the set $\mathit{FriendsOfFriends}$).
\end{itemize}

\begin{figure*}
\centering
\includegraphics[width=0.9\textwidth]{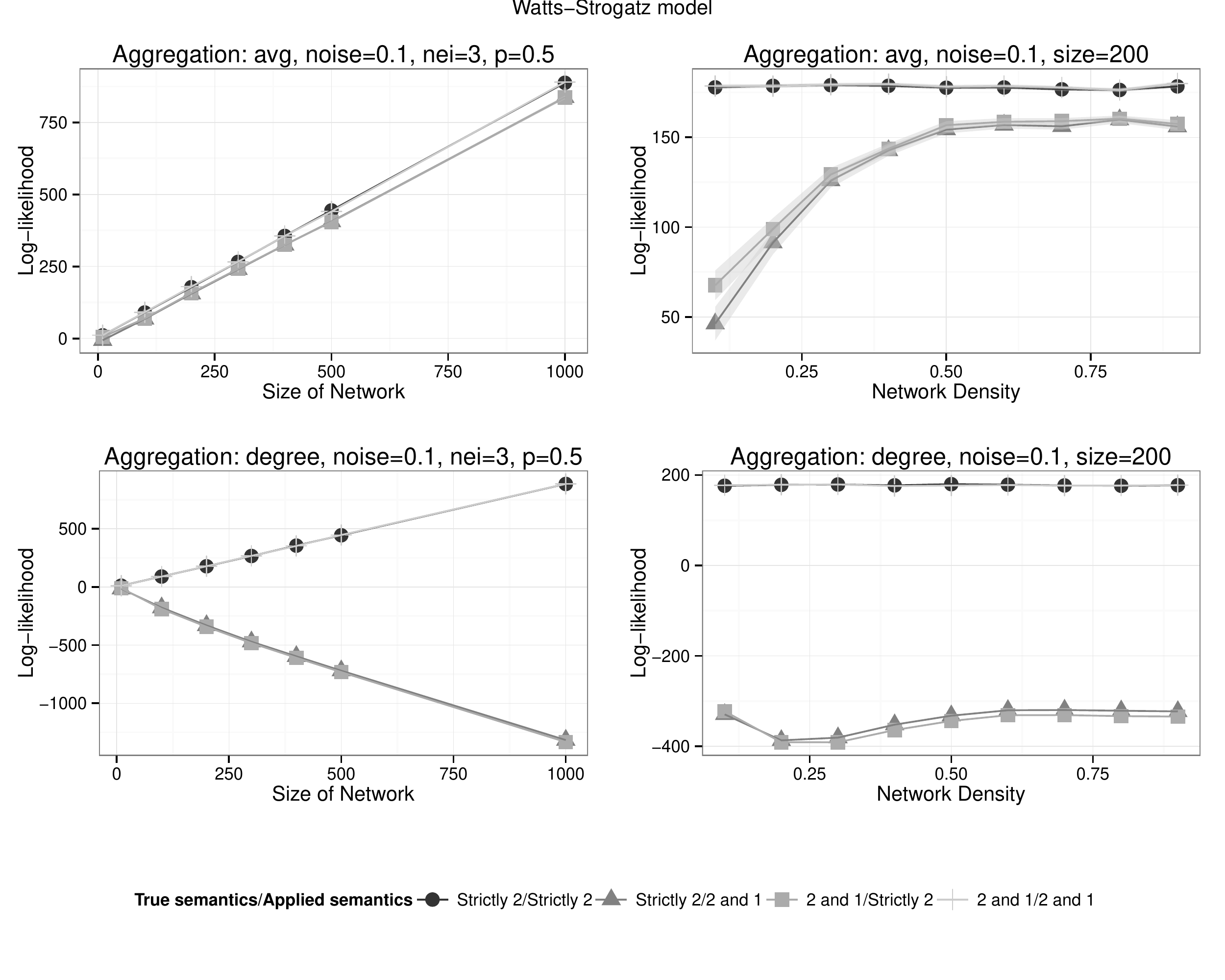}
\caption{Log-likelihood of model under varying network characteristics for graphs generated using the Erd{\"o}s-R\'{e}nyi model.}
\label{fig:erdos_normal_avg-degree}
\end{figure*}

\begin{figure*}
\centering
\includegraphics[width=0.9\textwidth]{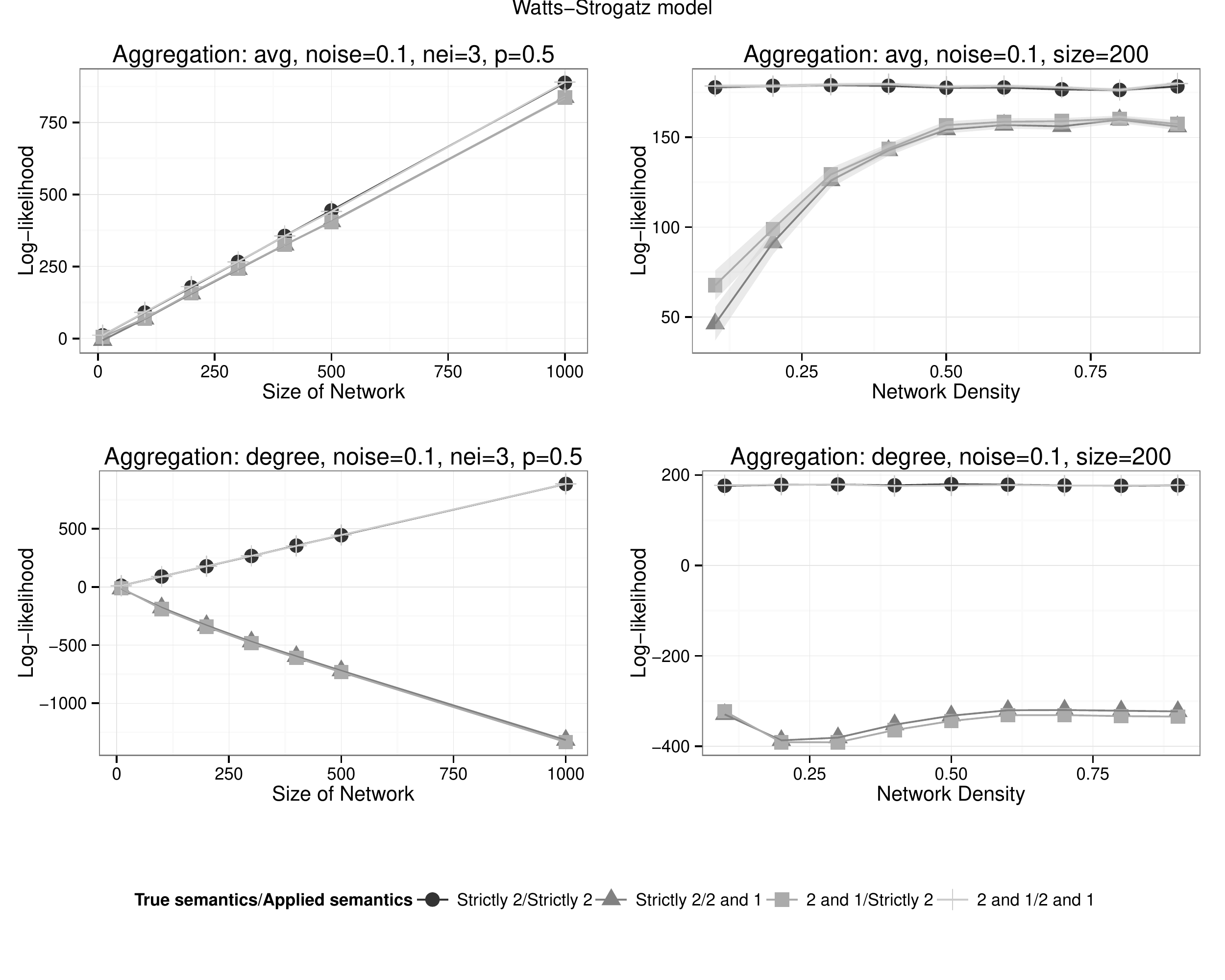}
\caption{Log-likelihood of model under varying network characteristics for networks generated using the Watts-Strogatz model.}
\label{fig:sw_normal_avg-degree}
\end{figure*}

We then learned two types of models from the generated data, one for each alternative definition of the set of friends of friends, and computed their log-likelihood.
The results for networks generated using the Erd{\"o}s-R\'{e}nyi model are shown in Figure~\ref{fig:erdos_normal_avg-degree} and for the Watts-Strogatz model in Figure~\ref{fig:sw_normal_avg-degree}.
We use the term ``strictly 2'' to refer to the semantics that include only friends of friends who are not the central node's friends, i.e., node with a shortest path of length 2 from the central node.
The term ``2 and 1'' refers to the semantics that would include friends in the set of friends of friends (nodes that have a path of length 2 from the central node, although the shortest path might be of length 1).
In general, when the generating semantics match the semantics assumed for the analysis, the model has a higher log-likelihood. 
Moreover, when comparing the two cases where the true generating semantics do not match the applied semantics, it is generally the case that the model that assumed that friends of friends include friends, performs better.
This is more pronounced for networks generated using the Erd{\"o}s-R\'{e}nyi model.

\section{Experiments on Real Networks}

\subsection{How different are the sets constructed using different semantics?}

To provide a more realistic insight on the impact of using different semantics, we measured the Jaccard distance between the sets constructed under different semantics on real network structures from the SNAP datasets~\cite{snapnets}.
In this case, there are no attribute values on the nodes.
The results are summarized in Table~\ref{tab:snap-results}.
These results indicate that there can be an average difference of up to $10\%$ on the sets constructed under different semantics.

\begin{table}
\caption{How different are the sets constructed using the two different semantics on some real-world network structures from the SNAP datesets.}
\label{tab:snap-results}
\centering
\tabcolsep=.08cm 
\begin{tabular}{l|cc|rrcc}
\multicolumn{1}{c}{} & \multicolumn{2}{c}{Experimental Results} &
\multicolumn{4}{c}{Network statistics}\\
\hline
 dataset & mean Jaccard & max Jaccard & Nodes & Edges & Diameter & Avg. Clustering Coeff.\\
 \hline
 as-733 &0.024 & 0.667 & 6,474 & 13,895 & 9 & 0.2522\\
 ca-AstroPh &0.013 & 1.0& 18,772 & 198,110 & 14 & 0.6306\\
 ca-CondMat &0.033 & 1.0& 23,133 & 93,497 & 14 & 0.6334\\
 ca-GrQc &0.046 & 1.0& 5,242 & 14,496 & 17 & 0.5296\\
 ca-HepPh &0.018 & 1.0& 12,008 & 118,521 & 13 & 0.6115\\
 ca-HepTh &0.024 & 1.0& 9,877 & 25,998 & 17 & 0.4714\\ 
 com-Amazon & 0.103 & 0.962 & 334,863 & 925,872 & 44 & 0.3967\\
 com-DBLP &0.096 & 0.933 & 317,080 & 1,049,866 & 21 & 0.6324\\
 email-Enron &0.068 & 1.0& 36,692 & 183,831 & 11 & 0.4970\\
 ego-Facebook &0.063 & 0.929 & 4,039 & 88,234 & 8 & 0.6055\\
 loc-Gowalla &0.040 & 0.875 & 196,591 & 950,327 & 14 & 0.2367\\
 Οregon-1 (010526) &0.001 & 0.4 & 11,174 & 23,409 & 9 & 0.2964\\
 Οregon-2 (010526) &0.001 & 0.524 & 11,461 & 32,730 & 9 & 0.4943\\
 roadNet-CA &0.047 & 1.0& 1,965,206 & 2,766,607 & 849 & 0.0464\\
 roadNet-PA &0.047 & 1.0& 1,088,092 & 1,541,898 & 786 & 0.0465\\
 roadNet-TX &0.047 & 1.0& 1,379,917 & 1,921,660 & 1054 & 0.0470\\
\hline
\end{tabular}
\end{table}

\section{Related Work}

There is a plethora of work analyzing the properties of social networks.
However, the actual construction for sets of peers is relatively understudied. 
A closely related work is that of Ugander et al.~\cite{ugander2011anatomy}, which focuses on analyzing the structure of the Facebook graph.
The authors explicitly make a distinction between the number of non-unique friends of friends (the number of paths of length two from the initial node that do not return to the initial node) and unique friends of friends (the set of nodes that can be reached from the initial node with paths of length two).
That distinction could be thought of as an additional type of semantics.
It is worth pointing out that our work has connections to feature construction for relational learning, as well as to probabilistic relational models, such as PRMs~\cite{GetoorSRL07} and Markov Logic Networks~\cite{richardson2006markov}.
More specifically, relational models are lifted representations that can be grounded to propositional models, according to some grounding semantics.
Those semantics specify how an edge on the relational level ``translates'' to an edge on the propositional level.

\section{Conclusions and Future Work}
The proliferation of network data presents an opportunity to study phenomena involving network effects such as social or peer effects at a scale previously unthinkable. 
While there has been a large amount of work in that area, there still exists a significant gap in terms of the exact semantics that can be used to describe the mechanism from which those phenomena arise. 
In this work we have looked at the semantics for constructing the set of peers in order to study a social or peer effect. 
While there is no definitive correct approach to construct these sets, the choice of strategy reflects assumptions about the underlying mechanism of influence. 
As both our synthetic and real-world experiments show, the difference is not merely theoretical; the use of one strategy over another can lead to significant differences in the aggregate values of network-based covariates. 
These results, when coupled with the importance of small-effect sizes in studies of large networks, demonstrate the necessity for clarity.
Our work here represents a small fraction of the ambiguities that exist within the process of understanding phenomena in networks. 

In future work, we plan to examine the effect of alternative semantics on causal estimates in network studies and to characterize the bias associated with inappropriately choosing the strategy for assembling sets. 
Another interesting line of inquiry is to extend our analysis to the case of multiple entity and multiple relation networks, such as networks that contain both \emph{friend} and \emph{co-worker} relations. 

\subsubsection*{Acknowledgments}
Funding was provided by the U.S. Army Research Office (ARO) and Defense Advanced Research Projects Agency (DARPA) under Contract Number W911NF-11-C-0088.  The content of the information in this document does not necessarily reflect the position or the policy of the Government, and no official endorsement should be inferred.  The U.S.\ Government is authorized to reproduce and distribute reprints for Government purposes notwithstanding any copyright notation here on.

\subsubsection*{References} 

\bibliographystyle{plainnat}
{\def\section*#1{}\bibliography{groundings}}

\end{document}